\documentclass[twocolumn,showpacs,showkeys,amssymb,nobibnotes,superscriptaddress,aps,pra,longbibliography]{revtex4-1}
\usepackage{graphicx}
\usepackage{dcolumn}
\usepackage{soul}
\usepackage{epsfig}
\usepackage{amsmath}
\usepackage{amsfonts}
\usepackage{dsfont}
\usepackage{amsmath,bm}
\usepackage{commath}
\usepackage{amssymb}
\usepackage{amssymb}
\usepackage{empheq}
\usepackage{xcolor}
\usepackage{bbm}
\usepackage{placeins}
\usepackage{soul}
\usepackage{physics}
\usepackage{times} 

\usepackage{setspace}
\usepackage{mathtools}
\usepackage{cases}
\usepackage{esint}
\usepackage{extarrows}
\usepackage{slashed}
\usepackage{pifont}
\usepackage[colorlinks,linkcolor=blue,urlcolor=blue,citecolor=blue]{hyperref}

\begin{document}
\author{Wen Huang}
\affiliation{School of Physics and Hubei Key Laboratory of Gravitation and Quantum Physics, Huazhong University of Science and Technology, Wuhan, 430074, P. R. China}
\author{Qian Bin}\email{qianbin@scu.edu.cn}
\affiliation{College of Physics, Sichuan University, Chengdu 610065, China}
\author{Ying Wu}
\affiliation{School of Physics and Hubei Key Laboratory of Gravitation and Quantum Physics, Huazhong University of Science and Technology, Wuhan, 430074, P. R. China}
\author{Xin-You L{\"u}}\email{xinyoulu@hust.edu.cn}
\affiliation{School of Physics and Hubei Key Laboratory of Gravitation and Quantum Physics, Huazhong University of Science and Technology, Wuhan, 430074, P. R. China}
	
\date{\today}
	
\title{Frequency-resolved $N$-photon correlations in the ultra-strong coupling regime}

\begin{abstract}
Frequency-resolved photon emission is central to applications from quantum information encoding to high-resolution spectroscopy, and then studying their correlations is therefore essential for revealing the underlying emission pathways and multiphoton statistics. Here, we investigate frequency-resolved $N$-photon correlations in an ultrastrongly coupled cavity QED system where a qubit interacts with a single-mode cavity. Owing to counter-rotating interactions, the eigenstates and energy spectrum are strongly modified, giving rise to rich spectral and statistical properties in the emitted frequency-resolved photons. Through frequency-selective detection, we reveal pronounced multiphoton antibunching, as well as multiphoton bunching originating from cascade transitions among dressed eigenstates. In particular, we show that parity symmetry plays a decisive role in shaping these correlations. The symmetry-breaking opens additional transition channels and dramatically enhances the generation of correlated photon pairs and even photon triplets of different frequencies. Our work extends  frequency-resolved correlations to the ultra-strong coupling regime and demonstrates their potential as a sensitive probe of symmetry in light–matter interaction systems.

		
\end{abstract}
	
\maketitle

\section{Introduction}\label{Sec I}
The quantum properties of light emitted from physical systems is a central topic in quantum optics. Not only does the emitted light field carry a lot of information about the system and its underlying dynamics, but it also has potential applications in the field of quantum information\,\cite{RMP2007,np2009,RMP2012Pan}. Photon correlation functions serve as a fundamental tool to characterize these quantum properties, with the second-order correlation function, $g^{(2)}(\tau)$, being the most commonly used. This quantity is typically measured using a Hanbury Brown-Twiss (HBT) interferometer in experiment\,\cite{HBT1956}. It characterizes the overall statistical features of the emitted light, enabling the clear distinction between photon bunching, photon antibunching, and coherent light\,\cite{Glauber1963prl,Glauber1963,Mandel1995}.
	
However, real emitted fields usually consist of various frequency components, as exemplified by the well-known "Mollow triplet"\,\cite{Mollow1969}. In such scenarios, conventional correlation measurement fails to resolve the intricate correlations between photons of different frequencies. With the development of quantum technology, the frequency-resolved photon correlation function $g^{(2)}(\omega_1, \omega_2; \tau)$ has been proposed\,\cite{Cohen1979,Dalibard1983,Knoll1986,SpectralCorrelations1993} and subsequently observed in experiments\,\cite{NP2012,PRB2015,Optica2018}. It specifically describes the correlation between a photon pair with frequencies $\omega_1$ and $\omega_2$. In order to efficiently compute $N$-photon correlations of any given frequencies, a general theoretical sensor method has been proposed\,\cite{PRL2012SensorMethod}.  Recently, promoted by its ability to reveal the hidden quantum phenomena\,\cite{pra2014nonclassical,PRB2015twocolorphoton, SciRep2016,prl2017FransonInterference},  frequency-resolved photon correlation has attracted widespread attention in various quantum systems\,\cite{KerrCavity2013,LPR2017,PRB2018Thirdorder,LiGaoXiang2019,pra2021twoatom,cavityoptomechanics2021,PRR2024TwoEmitter,prl2026Markovian}, including the cavity optomechanical system\,\cite{cavityoptomechanics2021}, coupled two-atom system\,\cite{LiGaoXiang2019,pra2021twoatom,PRR2024TwoEmitter}, and the Kerr-nonlinearity resonator\,\cite{KerrCavity2013}. However, current studies on frequency-resolved photon correlations have primarily focused on the weak- and strong-coupling regimes.
	
Ultrastrong coupling (USC) in cavity quantum electrodynamics (QED) represents a novel light-matter interaction regime\,\cite{nature2009USC,NPhy2010USC,PRB2009USC,pra2010NoclassicalState,ACS2014,NatPhys2017,NatureReviews2019USC,RMP2019USC,AQT2020} that surpasses the conventional weak and strong coupling regimes. In this regime, the coupling strength $g$ can reach or even exceed the order of $10\%$ of the cavity mode frequency (i.e., $g \gtrsim 0.1 \omega_c$). In recent years, USC has been experimentally realized on various platforms, such as superconducting quantum circuits\,\cite{NPhy2010USC,PRL2010superconducting,PRB2016TwomodeUSC,Bosman2017transmon}, semiconductor quantum wells\,\cite{PRL2009Todorov,PRL2012Semiconductor,Zhang2016NP,NatPhys2019ParaviciniBagliani} and other hybrid sysytems\,\cite{PRL2011MoleculeUSC,MEP2013USC,Zhang2014prlmagnon}. In the USC regime,  the rotating-wave approximation (RWA) is no longer valid. The presence of counter-rotating terms profoundly alters the properties of ground and excited states of the system, thereby giving rise to unexpected physical phenomena, including vacuum degeneracy\,\cite{PRL2010VacuumDegeneracy}, Casimir-like photons\,\cite{pra2011Casimir}, nonclassical states\,\cite{pra2010NoclassicalState}, and so on\,\cite{PRL2011PQC,PRL2013VirtualPhotons,prl2014Purcell,pra2018Franco,PRL2025Metrology,PRL2021Bin,pra2020Liao,PRL2018ShiTao,ChenYeHongPRL2024}. With experiments achieving the  continuously-increasing coupling strength, the interest in USC systems continues to grow. On the one hand, significant efforts have been devoted to establishing complete theoretical frameworks\,\cite{PRA2018Rabl,NatPhys2019GaugeAmbiguities,PRR2021Gauge,PRR2023Gauge,PRL2024MasterEquation}, such as gauge invariance in USC cavity QED\,\cite{PRA2018Rabl,NatPhys2019GaugeAmbiguities,PRR2021Gauge,PRR2023Gauge}. On the other hand, extensive studies have investigated the unique influence of USC on various quantum effects\,\cite{PRL2012PhotonBlockade,PRA2016PB,pra2016PhaseTransition,PRA2017DickeStates,PRB2020LandauPolaritons,PRL2021Nonlinearities,PRL2022RamanPhotons,pra2022PhotonPhononBlockade,PRL2023Squeezing,PRA2023NonGaussianSuperradiant,PRL2026TPB}. Given the complex dressed-state structure inherent to USC systems, the quantum statistics of their emitted light are expected to be interesting. Although some studies have explored the statistical behaviors of the emitted light, the correlation between photons of different frequencies in the USC regime remains unexplored.

Here, we theoretically investigate frequency-resolved photon correlations in the USC regime, with particular emphasis on the role of parity symmetry. Using the sensor approach, we efficiently calculate the power spectrum as well as frequency-resolved second- and three-order correlation functions. The power spectrum exhibits a multi-peak structure, indicating that the emitted light contains photons with multiple frequency components. The correlations among photons of different frequencies display remarkably rich behaviors. For two-photon processes, we observe both frequency-resolved photon bunching and antibunching. Extending the analysis to high-order process reveals even stronger frequency-resolved three-photon bunching effects. These correlations can be understood within the dressed-state picture, where the underlying mechanism originates from cascade transitions between the eigenstates of the  system. Importantly, we find that parity symmetry plays a crucial role in shaping photon correlations. Breaking parity symmetry opens transition channels that are otherwise forbidden, thereby significantly enhancing the generation of correlated photon pairs and photon triplets with different frequencies. Our work provides a theoretical foundation for the generation of multi-photon sources, and has potential applications in quantum information and quantum metrology.
	
\section{MODEL}\label{sec II}
		
\begin{figure}
	\centering
	\includegraphics[width=1.0\linewidth]{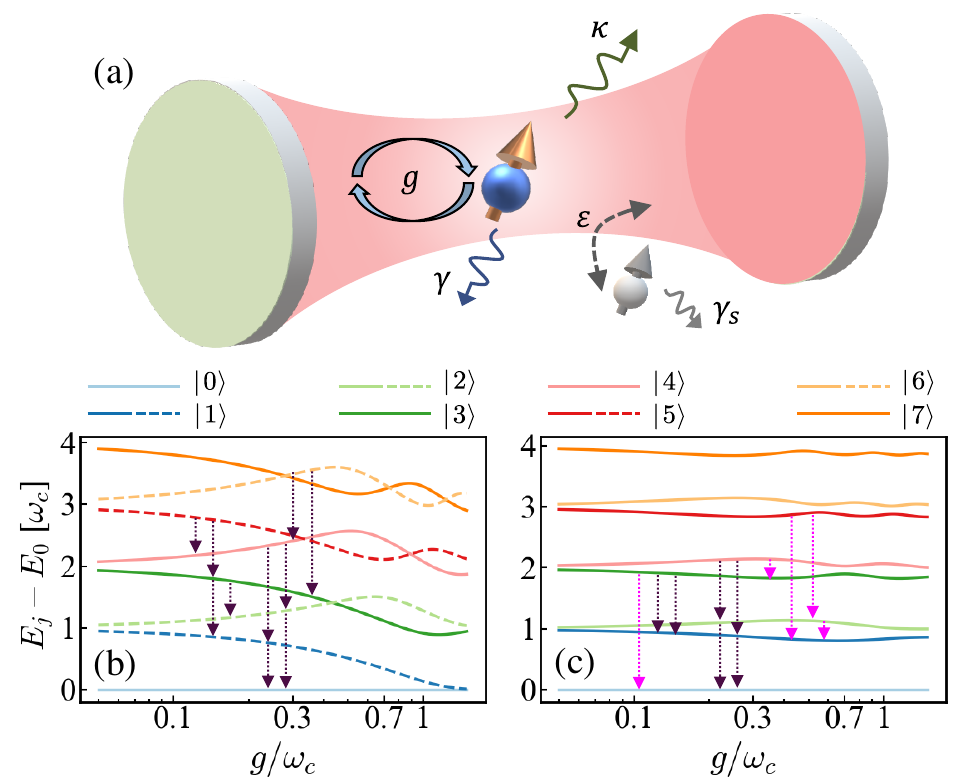}
	\caption{(a) Schematic setup of the cavity QED system: a system qubit ultrastrongly interacts with a cavity mode with strength $g$, and a sensor qubit is weakly couples to the cavity with a vanishing coupling strength $\varepsilon$. The symbols $\kappa$ and $\gamma$ ($\gamma_s$) represent the decay rates of the cavity and the system qubit (sensor qubit), respectively. (b),(c) Energy Spectrum of $\widetilde{H}_{\text{QR}}$ as a function of the coupling strength $g$ for (b) $\theta=\pi/2$ and (c) $\theta=\pi/6$.  In panel (b), the solid and dashed lines indicate energy levels corresponding to the eigenstates with even and odd parity, respectively. In panel (c), the lines simply indicate the energy levels and do not correspond to parity. The arrows illustrate the radiative decay transitions between eigenstates discussed later. Especially, the magenta arrows denote the symmetry-breaking-induced transitions which are forbidden in the symmetry-preserving case. In panels (b) and (c), the parameters are choose as $\omega_q = \omega_c$.}
	\label{fig:model and energy spectrum}
\end{figure}

We consider a cavity QED system, as shown in Fig.\,\ref{fig:model and energy spectrum}(a), consisting of a single-mode cavity field ultrastrongly coupled to a system qubit  (the blue one). To investigate the frequency-resolved photon correlations of the emitted photons from the system, we employ the sensor method based on perturbation theory\,\cite{pra2018PerturbationSensor}. In this approach, auxiliary sensor qubits are weakly coupled to the cavity and act as frequency filters. To avoid perturbing the system dynamics, the coupling between the sensor qubit and the cavity is assumed to be sufficiently weak. This method allows us to efficiently compute the correlations of frequency-filtered photons directly by  population correlations of the sensor qubits, without evaluating complicated multidimensional integrals.


In the dipole gauge, the total Hamiltonian of the system including the sensor qubit takes the form ($\hbar=1$)
\begin{eqnarray}\label{eq:total Hamiltonian}
H=\widetilde{H}_{\text{QR}} + H_{\text{1}} + H_{s,1}.
\end{eqnarray}
Here, $\widetilde{H}_{\text{QR}}$ denotes an extended quantum Rabi Hamiltonian describing the interaction between the system qubit and the cavity field, 
\begin{eqnarray}\label{eq:Rabi model}
\widetilde{H}_{\text{QR}}=\omega_c a^{\dagger} a + \omega_q \sigma_{+}\sigma_{-}-i g \left(a-a^{\dagger}\right)\sigma_p,
\end{eqnarray}
where $a$ ($a^{\dagger}$) is the annihilation (creation) operator of the cavity field with frequency $\omega_c$, $\sigma_{x}=\sigma_{+}+\sigma_{-}$ ($\sigma_{+}=|e \rangle \langle g |$, $\sigma_{+}=|g \rangle \langle e |$) and $\sigma_z$ are Pauli operators of the system qubit with frequency $\omega_q$, and the operator $\sigma_p$ is defined as $\sigma_p=\cos \theta \sigma_z-\sin \theta \sigma_x$. The parameter $g$ quantifies the overall coupling strength between the system qubit and the cavity, while the angle $\theta$ determines the relative contributions of the transverse and longitudinal couplings. Importantly, $\theta$ also governs the parity symmetry of the Hamiltonian $\widetilde{H}_{\text{QR}}$. For $\theta=m \pi/2$ ($m$ being an odd number), $\widetilde{H}_{\text{QR}}$ reduces to the standard quantum Rabi Hamiltonian $H_{QR}$, which possesses a parity symmetry. The corresponding parity operator $\Pi=e^{i \pi\left[a^{\dagger} a+\frac{1}{2}\left(1+\sigma_z\right) \right]}$, commutes with $H_{QR}$, indicating that the parity of the total number of excitations is conserved. Consequently, the eigenstates can be classified according to their odd or even parity. When $\theta\neq m \pi/2$, however, this symmetry is broken. 

Figures\,\ref{fig:model and energy spectrum} (b,c) show the first eight eigenenergies and the allowed transitions between eigenstates of the Hamiltonian $\widetilde{H}_{\text{QR}}$ for $\theta=\pi/2$ and $\theta=\pi/6$, respectively. In Fig.\,\ref{fig:model and energy spectrum} (b), where parity symmetry is preserved, the energy levels  are classified by the parity of the their corresponding eigenstates: solid and dashed lines indicate even- and odd-parity eigenstates, respectively. In Fig.\,\ref{fig:model and energy spectrum} (c), the lines simply represent energy levels and no longer correspond to parity. The arrows indicate the possible radiative transitions between eigenstates that contribute to the power spectra shown later in Fig.\,\ref{fig:spectrum}. Since our analysis does not involve transitions to higher excited states (such as levels 7 and 8), the corresponding transitions are not shown in Fig.\,\ref{fig:model and energy spectrum}(c).

	
The second term of the total Hamiltonian describes the free Hamiltonian of the sensor qubit, $H_1 =\omega_1 \varsigma_{1}^{+}\varsigma_{1}^{-}$, while the last term describes the interaction Hamiltonian between the system and the sensor,
\begin{eqnarray}\label{eq:sensor Hamiltonian}
H_{\varsigma,1}=\varepsilon [i(a^{\dagger}-a)+2\eta \sigma_x]\varsigma_{1}^{x},
\end{eqnarray}
where $\varsigma_{1}^{x}=\varsigma_{1}^{+}+\varsigma_{1}^{-}$, $\varsigma_{1}^{\pm}$ are Pauli operators of the sensor qubit with frequency $\omega_1$, and $\eta=g/\omega_c$ is the normalized coupling parameter. The coupling strength $\varepsilon$ between the sensor and the main system must be small enough so that the sensor does not significantly modify the system dynamics. For  simplicity, the schematic in Fig.\,\ref{fig:model and energy spectrum}(a) illustrates only a single sensor qubit. In general, however, the method can be extended to $N$ sensors, where the number of sensors is corresponds to the order $N$ of the photon correlations to be calculated.
	
To describe a real system, its interaction with the external environment must be considered. In general, the dynamics of an open quantum system in the weak or strong coupling regime can be described using a standard quantum master equation. However, such an approach leads to unphysical predictions in the USC regime ($g/\omega_c \gtrsim 0.1$). A typical example is that the system cannot be relaxed to its ground state even in a zero-temperature environment. Therefore, in the USC regime the light–matter system must be treated as an indivisible entity, and its interaction with the environment should be described in the basis formed by the dressed states, i.e., the eigenstates of the full system Hamiltonian. One can then derive a generalized master equation in this dressed-state basis to describe the dissipative processes \,\cite{PRL2012PhotonBlockade}. In this framework, the dissipators describe transitions between dressed states rather than the decay of individual photons or atoms.
	
In the limit of weak system-sensor coupling strength, it is sufficient to express the system Hamiltonian in terms of the eigenstates $\vert j \rangle$ ($j=0,1,2, \cdots$) of $\widetilde{H}_{\text{QR}}$ with eigenvalues $E_j$, i.e., $\widetilde{H}_{\text{QR}} \vert j \rangle=E_j \vert j \rangle$. The eigenstates are ordered such that $E_k>E_j$ for $k>j$. Under weak incoherent driving and in a zero-temperature environment, the generalized master equation takes the form
\begin{equation}\label{eq:Master_equation}
\frac{d \rho}{d t} = \mathcal{L}\rho = \mathcal{L}_{0}\rho -i[H_{s,1},\rho] + \mathcal{L}_{1}\rho,
\end{equation}
where
\begin{align}\label{eq:Master_equation_system}
\mathcal{L}_{0}\rho=& i[\widetilde{H}_{\text{QR}}, \rho] + \kappa \mathcal{L}\left[X^{+}\right]\rho + \gamma \mathcal{L}\left[D^{+}\right]\rho \\ \nonumber
& + P_{\text{inc}} \mathcal{L}\left[X^{-}\right]\rho
\end{align}
and
\begin{align}\label{eq:Master_equation_sensor}
\mathcal{L}_{1}\rho=-i[H_{1},\rho] + \Gamma \mathcal{L}\left[\varsigma_{1}^{-}\right]\rho.
\end{align}
Here the Liouvillian superoperator is defined as $\mathcal{L}[O]=O\rho O^{\dagger}-\frac{1}{2}O^{\dagger}O\rho-\frac{1}{2}\rho O^{\dagger}O$, and $\rho$ is the composite system-sensor density matrix. The parameters $\kappa$, $\gamma$, and $\Gamma$ represent the decay rates of the cavity, the system qubit, and sensor qubit, respectively, while $P_{\text{inc}}$ describes the incoherent pumping strength. The dissipation operators are defined as 
\begin{equation}\label{eq:dissipation operators}
\begin{aligned}
&X^{+}=\sum_{j, k>j}\langle j \vert i [(a^{\dagger}-a)+2\eta \sigma_x]  \left|k\right\rangle \left|j\right\rangle\left\langle k\right| \omega_{kj}/\omega_c,\\
&D^{+}=i\sum_{j, k>j}\langle j \vert \sigma_x \left|k\right\rangle \left|j\right\rangle\left\langle k\right| \omega_{kj}/\omega_q,
\end{aligned}
\end{equation}
where $X^{+}$ and  $D^{+}$ correspond to the positive-frequency components of cavity field and system qubit, respectively, and $\omega_{kj}=\omega_k-\omega_j>0$. These operators satisfy the relations $X^{-}=[X^{+}]^{\dagger}$ and $D^{-}=[D^{+}]^{\dagger}$.

\section{Power spectrum in the ultrastrong coupling regime }\label{sec III}
	
According to the sensor method\,\cite{PRL2012SensorMethod}, the power spectra of emission can be defined as
\begin{equation}\label{eq:spectrum_1}
S(\omega_1)=\lim _{\varepsilon \rightarrow 0} \frac{\Gamma}{2 \pi \varepsilon^2}\left\langle\varsigma_1^{\dagger} \varsigma_1\right\rangle,
\end{equation}
where $\langle\varsigma_1^{\dagger} \varsigma_1\rangle$ is the occupation of sensor qubit with transition frequency $\omega_1$ and decay rate $\Gamma$. To ensure the validity of this method, the system-sensor coupling strength $\varepsilon$ must satisfy $\varepsilon \ll \sqrt{\Gamma \gamma_Q/2}$, where $\gamma_Q$ is the smallest decay rate in the system. This approach is conceptually straightforward, as it only requires incorporating the sensor into the system dynamics. However, the computation becomes numerically intensive or even prohibitive for systems with large Hilbert spaces, especially when higher-order photon correlations are considered. Moreover, the choice of sensor-system coupling $\varepsilon$ involves a critical trade-off: it must be sufficiently small to avoid perturbing the system dynamics, while not being excessively small so as to prevent numerical precision issues when calculating correlations. To overcome these difficulties, an alternative perturbative sensor formulation was proposed\,\cite{pra2018PerturbationSensor}. This method is based on an algebraic expansion of the composite steady state (including the sensor) in powers of the coupling parameter $\varepsilon$, which can then be perturbatively truncated to evaluate the physical quantities of interest.

\begin{figure}
	\centering
	\includegraphics[width=1.0\linewidth]{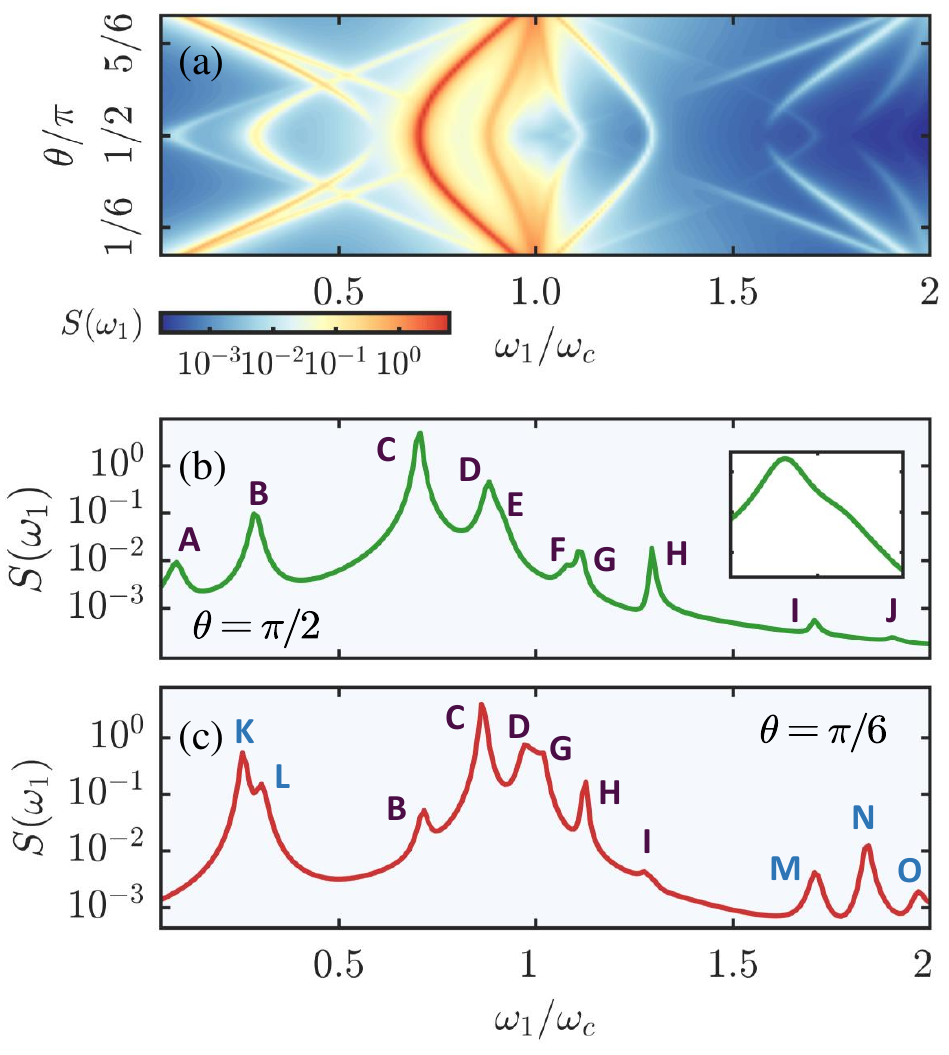}
	\caption{Power spectra of emission $S(\omega_1)$ under weak incoherent pumping. (a) Power spectrum versus the angle $\theta$ and the sensor scanning frequency $\omega_1$. (b),(c) Line cuts of $S(\omega_1)$ for (b) $\theta=\pi/2$ and (c) $\theta=\pi/6$. The labeled peaks correspond to transitions between different eigenstates, summarized in Table\,\ref{tab:tab1}. The labels are color-coded to distinguish the different types of transitions: deep purple denotes conventional transitions between eigenstates with opposite parity, whereas blue denotes transitions induced by symmetry breaking. The inset in panel (b) shows an enlarged view of the region around peaks $\text{D}$ and $\text{E}$, highlighting the subtle peak $\text{E}$. Parameters are $\omega_q/\omega_c=1$, $g/\omega_c=0.3$, $\varepsilon/\omega_c=10^{-4}$, $\kappa/\omega_c=5\times10^{-3}$, $\gamma=\kappa$, and $\Gamma=\kappa$, $P_{\text{inc}}=0.1\kappa$.}
	\label{fig:spectrum}
\end{figure}
In the following, we apply this method to calculate the power spectrum in the ultrastrong coupling regime. Firstly, we insert the identity operator $\mathbb{I}_{\varsigma_1}=\sum_{m=0,1}\left|m\right\rangle\left\langle m\right|$, defined in the sensor qubit Hilbert space, into the joint system-sensor steady state $\rho_{ss}$. The joint steady state can be rewritten as
\begin{equation}
\rho_{ss}=\mathbb{I}_{\varsigma_1} \rho_{s s} \mathbb{I}_{\varsigma_1}=\sum_{m, m^{\prime}=0,1} \rho_{m}^{m^{\prime}}\otimes\left|m\right\rangle\left\langle m^{\prime}\right|,
\end{equation}
where the matrices $\rho_{m}^{m^{\prime}}=\langle m | \rho_{s s} |m^{\prime}\rangle$ ($\rho_{m^{\prime}}^{m}={\rho_{m}^{m^{\prime}}}^{\dagger}$) are the so-called ``auxiliary conditional states''. There matrices are defined on the system Hilbert space but are conditioned on the state of the sensor qubit. Within this representation, the power spectra in Eq.\,\eqref{eq:spectrum_1} takes the form
\begin{equation}\label{eq:spectrum_2}
S(\omega_1)=\frac{\Gamma}{2 \pi \varepsilon^2}\text{Tr}[\rho_{1}^{1}].
\end{equation}
Note that the restriction on $\varepsilon$ is omitted here, as it only appears as a global scaling factor and does not affect the spectral profile. Solving the steady state $\mathcal{L}\rho_{s s}=0$ and following the procedure described in Ref.\,\cite{pra2018PerturbationSensor}, the auxiliary conditional states can be obtained as
\begin{subequations}
\label{eq:ACS} 
\begin{align}
|\rho_1^0\rangle\rangle & \sim \frac{i\varepsilon X^{'} |\rho_0^0\rangle\rangle}{\mathcal{L}_0 - (\Gamma/2 + i\omega_1)\mathbb{I}}, \label{eq:a} \\
|\rho_1^1\rangle\rangle & = \frac{i\varepsilon(X^{'} |\rho_0^1\rangle\rangle - |\rho_1^0\rangle\rangle X^{'\dagger})}{\mathcal{L}_0 - \Gamma\mathbb{I}}, \label{eq:b}
\end{align}
\end{subequations}
where
\begin{equation}
X^{'}=\sum_{j, k>j}\langle j \vert i [(a^{\dagger}-a)+2\eta \sigma_x]  \left|k\right\rangle \left|j\right\rangle\left\langle k\right|.
\end{equation}
Equation\,\eqref{eq:ACS} is expressed in Liouville space, in which density matrices are treated as vectors, and operators are treated as superoperators. The corresponding mappings are
\begin{align}
X^{'} |\rho_m^{m^{\prime}}\rangle\rangle & \rightarrow (X^{'} \otimes \mathbb{I}) |\rho_m^{m^{\prime}}\rangle\rangle,  \\ \nonumber
|\rho_{m}^{m^{\prime}}\rangle\rangle X^{'\dagger} & \rightarrow (\mathbb{I}  \otimes (X^{'\dagger})^{T}) |\rho_m^{m^{\prime}}\rangle\rangle, 
\end{align}
where an $N \times N$ density matrix $\rho_m^{n}$ ($N$ is the dimension of the system Hilbert space)  is mapped onto an $N^2$-dimendional vector $|\rho_m^{m^{\prime}}\rangle\rangle$, while an $N \times N$ operator $X^{'}$ becomes $N^2 \times N^2$ superoperator, and $\mathbb{I}$ is the identity matrix in the system Hilbert space. Consequently, obtaining $\rho_{1}^{1}$ only requires  solving the steady state $\rho_{0}^{0}$, which is independent of the sensor. Substituting Eq.\,\eqref{eq:ACS} into Eq.\,\eqref{eq:spectrum_2}, one obtains a semi-analytical solution for the power spectrum, in which the dependence on $\varepsilon$ vanishes.

Figure\,\ref{fig:spectrum} shows the power spectrum $S(\omega_1)$ as a function of the angle $\theta$ under weak incoherent driving, where the system's eigenstates are not significantly perturbed. As shown in Fig.\,\ref{fig:spectrum}(a), the spectral peaks trace out characteristic trajectories as $\theta$ varies. The peak positions correspond to radiative transition frequencies between the system eigenstates. The transition $|1\rangle \rightarrow |0\rangle$ consistently exhibits the strongest spectral peak for all values of $\theta$, as it corresponds to the dominant radiative decay channel from the lowest excited state, which typically carries the largest steady-state population. Most importantly, along the horizontal line $\theta=\pi/2$, where the system possesses parity symmetry, several spectral peaks are strongly suppressed or vanish entirely. This behavior reveals the profound influence of parity symmetry on the emission spectrum. As discussed earlier, the parameter $\theta$ controls the parity symmetry of the Hamiltonian, which determines whether transitions between eigenstates within the same parity subspace are allowed.

	
%
\begin{table}
\renewcommand{\arraystretch}{0.8}
\centering
\caption{Assignment of spectral peaks to the corresponding transitions in Fig.\,\ref{fig:spectrum}(b,c), where $k \rightarrow j$ ($k,j=0,1,2,...$ and $k>j$  ) represents the transition from state $\vert k \rangle$ to state $\vert j \rangle$.}
\label{tab:tab1}
\begin{center}
\begin{tabular}{ccccc}
\toprule 
\text{Peak} \quad & \text{Transition}\quad &\quad\quad\quad & \text{Peak} \quad & \text{Transition}\quad \\
\hline
A & $5\rightarrow4$ &\quad\quad\quad & I & $4\rightarrow1$ \\
B & $3\rightarrow2$ &\quad\quad\quad & J & $7\rightarrow3$ \\
C & $1\rightarrow0$ &\quad\quad\quad & K & $2\rightarrow1$ \\
D & $3\rightarrow1$ &\quad\quad\quad & L & $4\rightarrow3$ \\
E & $5\rightarrow3$ &\quad\quad\quad & M & $5\rightarrow2$ \\
F & $7\rightarrow4$ &\quad\quad\quad & N & $3\rightarrow0$ \\
G & $4\rightarrow2$ &\quad\quad\quad & O & $5\rightarrow1$ \\
H & $2\rightarrow0$ &\quad\quad\quad &   &                 \\
\hline\hline
\end{tabular}
\end{center}
\end{table}
\begin{figure}[htbp]
	\centering
	\includegraphics[width=1.0\linewidth]{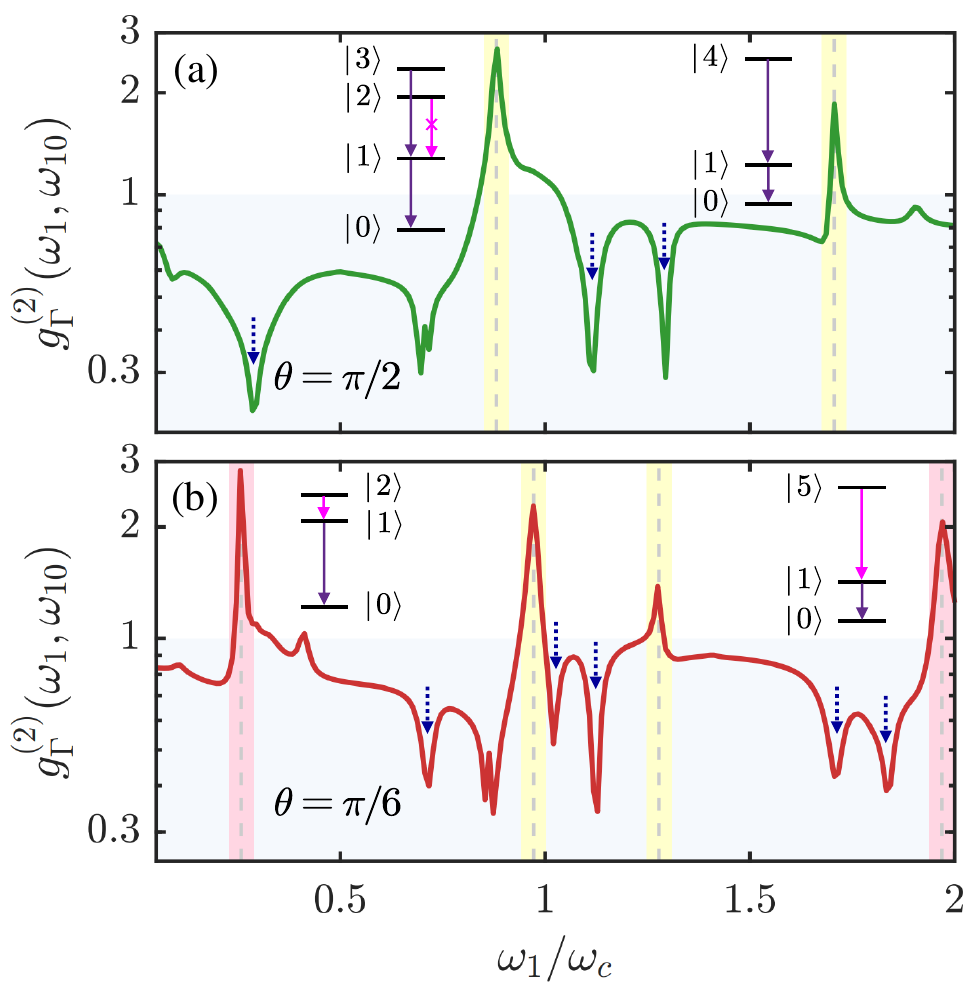}
	\caption{Second-order photon correlation functions $g_{\Gamma}^{(2)}(\omega_1, \omega_{10})$ vs sensor resonance frequency $\omega_1$ for (a) $\theta=\pi/2$ and (b) $\theta=\pi/6$, when one sensor is fixed at the fundamental transition frequency $\omega_{10}$. The light-blue background represents the antibunching regime ($g_{\Gamma}^{(2)}(\omega_1, \omega_{10})<1$), whereas the peaks (shaded in yellow and pink) indicate photon bunching ($g_{\Gamma}^{(2)}(\omega_1, \omega_{10})>1$). Several key dips are marked by blue dotted arrows. A key feature is the emergence of additional bunching peaks (pink) in panel (b), which are absent in panel (a). The parameters are the same as in Fig.\,\ref{fig:spectrum}.}
	\label{fig:twophoton}
\end{figure}
To illustrate the spectral features more clearly, we choose $\theta=\pi/2$ and $\theta=\pi/6$ and plot the corresponding spectra in Figs.\,\ref{fig:spectrum}(b) and (c), respectively. For clarity in the subsequent discussion, the observed spectral peaks are labeled, and their corresponding transition assignments are summarized in Table\,\ref{tab:tab1}. As shown in Fig.\,\ref{fig:spectrum}(b), the spectrum exhibits eight distinct peaks (corresponding to the arrows in Fig.\,\ref{fig:model and energy spectrum}(b)). Among them, the most prominent peak B corresponds to the transition $|1\rangle \rightarrow |0\rangle$. In addition, a subtle but genuine peak (labeled E) is also present, as highlighted in the inset of Fig.\,\ref{fig:model and energy spectrum}(b). This peak originates from the transition $|5\rangle \rightarrow |3\rangle$, which plays an important role in the frequency-resolved three-photon correlation discussed latter. As is shown in Fig.\,\ref{fig:model and energy spectrum}(b), the transitions indicated by arrows occur exclusively between energy levels of opposite parity (solid and dashed lines), whereas transitions between states with the same parity are forbidden. This behavior follows directly from symmetry-determined selection rules. When $\theta=\pi/2$, the parity of the system is conserved. Since the operator responsible for transitions $a-a^\dagger$ (or $\sigma_x$) in Eq.\,\eqref{eq:dissipation operators} changes the parity of the state, transitions between eigenstates within the same parity subspace are forbidden. For $\theta=\pi/6$, however, the parity symmetry of the system is broken due to the presence of the $(a-a^{\dagger})\sigma_z$ term. Consequently, transitions that are forbidden at $\theta = \pi/2$ can become allowed when $\theta = \pi/6$. As depicted in Fig.\,\ref{fig:spectrum}(c), the spectrum becomes significantly more complex, with several additional peaks appearing that are absent in Fig.\,\ref{fig:spectrum}(b). 

According to Table\,\ref{tab:tab1}, peaks $\{ \text{K, L, M, N, O} \}$, originating from the transitions $ \{\vert 2 \rangle \rightarrow \vert 1 \rangle, \vert 4 \rangle \rightarrow \vert 3 \rangle, \vert 5 \rangle \rightarrow \vert 2 \rangle, \vert 3 \rangle \rightarrow \vert 0 \rangle, \vert 5 \rangle \rightarrow \vert 1 \rangle \}$, correspond to transitions between states that share the same parity in the energy spectrum at $\theta=\pi/2$. 
For instance, the states $\vert 2 \rangle$ and $\vert 1 \rangle$ involved in the transition $\vert 2 \rangle \rightarrow \vert 1 \rangle$ both have odd parity, whereas states $\vert 3 \rangle$ and $\vert 0 \rangle$ both possess even parity. These additional spectral peaks  arise from the transitions induced by the symmetry-breaking of the system Hamiltonian, which are indicated by the magenta arrows in Fig.\,\ref{fig:model and energy spectrum}(c). The emergence of these new transition channels not only produces additional spectral peaks but also redistributes the spectral weight, suppressing the transitions involving higher excited states that are responsible for peaks $\text{E}$ and $\text{F}$.
This spectral redistribution has a significant impact on frequency-resolved photon correlations, which will be analyzed detailedly in the next section. 

\section{ Frequency-resolved two-photon and three-photon correlations}\label{sec IV}

While the power spectrum reveals the frequencies of the photons emitted by the system, it does not provide information about correlations between photons of different frequencies. In the case of zero delay time, the normalized second-order photon correlation obtained from the intensity–intensity cross-correlations between two sensors is given by
\begin{equation}
g_{\Gamma}^{(2)}(\omega_1, \omega_2)=\lim _{\varepsilon \rightarrow 0}\frac{ \left\langle\varsigma_1^{\dagger} \varsigma_1\varsigma_2^{\dagger} \varsigma_2\right\rangle}{\left\langle\varsigma_1^{\dagger} \varsigma_1 \right\rangle \left\langle\varsigma_2^{\dagger} \varsigma_2\right\rangle},
\end{equation}
where $\varsigma_2^{\dagger}\varsigma_2$ is the occupation of the second sensor with frequency $\omega_2$. The second sensor has the same coupling form and dissipation rate as the first, and all sensors introduced in this study are treated as identical. Applying the same procedure as before, the second-order photon correlation can be expressed in terms of the matrices $\tilde{\rho}_{m,n}^{m^{\prime},n^{\prime}}$, 
\begin{equation}
g_{\Gamma}^{(2)}(\omega_1, \omega_2)=\frac{\text{Tr}[\tilde{\rho}_{1,1}^{1,1}]}{\text{Tr}[\tilde{\rho}_{1,0}^{1,0}]\text{Tr}[\tilde{\rho}_{0,1}^{0,1}]},
\end{equation}
where $\tilde{\rho}_{m,n}^{m^{\prime},n^{\prime}}=\rho_{m,n}^{m^{\prime},n^{\prime}}/\varepsilon^{m+m^{\prime}+n+n^{\prime}}$ with the ``auxiliary conditional states'' $\rho_{m,n}^{m^{\prime},n^{\prime}}=\langle m,n |\rho_{ss}|m^{\prime}, n^{\prime}\rangle$ in the case of two sensors. Following the procedure analogy to Eq.\,\eqref{eq:ACS}, the solutions of the matrices read
\begin{subequations}
\label{eq:ACS1} 
\begin{align}
|\tilde{\rho}_{1,0}^{0,0}\rangle\rangle & \sim \frac{i X^{'} |\tilde{\rho}_{0,0}^{0,0}\rangle\rangle}{\mathcal{L}_0 - ( i\omega_1 +\Gamma/2)\mathbb{I}}, \label{eq:A} \\
|\tilde{\rho}_{0,1}^{0,0}\rangle\rangle & \sim \frac{i X^{'} |\tilde{\rho}_{0,0}^{0,0}\rangle\rangle}{\mathcal{L}_0 - ( i\omega_2 +\Gamma/2)\mathbb{I}}, \label{eq:B} \\
|\tilde{\rho}_{1,0}^{1,0}\rangle\rangle & \sim \frac{i(X^{'}|\tilde{\rho}_{0,0}^{1,0}\rangle\rangle - |\tilde{\rho}_{1,0}^{0,0}\rangle\rangle X^{'\dagger})}{\mathcal{L}_0-\Gamma \mathbb{I}}, \label{eq:C} \\
|\tilde{\rho}_{0,1}^{0,1}\rangle\rangle & \sim \frac{i(X^{'}|\tilde{\rho}_{0,0}^{0,1}\rangle\rangle - |\tilde{\rho}_{0,1}^{0,0}\rangle\rangle X^{'\dagger})}{\mathcal{L}_0-\Gamma \mathbb{I}}, \label{eq:D} \\
|\tilde{\rho}_{1,1}^{0,0}\rangle\rangle & \sim \frac{i(X^{'}|\tilde{\rho}_{0,1}^{0,0}\rangle\rangle + X^{'}|\tilde{\rho}_{1,0}^{0,0}\rangle\rangle )}{\mathcal{L}_0-(i\omega_1 +i\omega_2+\Gamma) \mathbb{I}}, \label{eq:E} \\
|\tilde{\rho}_{1,0}^{0,1}\rangle\rangle & \sim \frac{i(X^{'}|\tilde{\rho}_{0,0}^{0,1}\rangle\rangle - |\tilde{\rho}_{1,0}^{0,0}\rangle\rangle X^{'\dagger})}{\mathcal{L}_0-(i\omega_1 - i\omega_2+\Gamma) \mathbb{I}}, \label{eq:F} \\
|\tilde{\rho}_{1,1}^{0,1}\rangle\rangle & \sim \frac{i(X^{'}|\tilde{\rho}_{0,1}^{0,1}\rangle\rangle + X^{'}|\tilde{\rho}_{1,0}^{0,1}\rangle\rangle - |\tilde{\rho}_{1,1}^{0,0}\rangle\rangle X^{'\dagger})}{\mathcal{L}_0-(i\omega_1 +3\Gamma/2) \mathbb{I}}, \label{eq:G} \\
|\tilde{\rho}_{1,1}^{1,0}\rangle\rangle & \sim \frac{i(X^{'}|\tilde{\rho}_{0,1}^{1,0}\rangle\rangle - |\tilde{\rho}_{1,1}^{0,0}\rangle\rangle X^{'\dagger} + X^{'}|\tilde{\rho}_{1,0}^{1,0}\rangle\rangle)}{\mathcal{L}_0-(i\omega_2 +3\Gamma/2) \mathbb{I}}, \label{eq:H} \\
|\tilde{\rho}_{1,1}^{1,1}\rangle\rangle & = \frac{i(X^{'}|\tilde{\rho}_{0,1}^{1,1}\rangle\rangle - |\tilde{\rho}_{1,1}^{0,1}\rangle\rangle X^{'\dagger} + X^{'}|\tilde{\rho}_{1,0}^{1,1}\rangle\rangle - |\tilde{\rho}_{1,1}^{1,0}\rangle\rangle X^{'\dagger})}{\mathcal{L}_0-2\Gamma \mathbb{I}}. \label{eq:I}
\end{align}
\end{subequations}
This calculation can be straightforwardly extended to third- and even higher-order photon correlations. However, the resulting semi-analytical expressions become exceedingly lengthy and cumbersome. For brevity, we therefore do not present the explicit form of the third-order correlation function $g_{\Gamma}^{(3)}(\omega_1, \omega_2, \omega_3)$.

\begin{figure}
	\centering
	\includegraphics[width=1.0\linewidth]{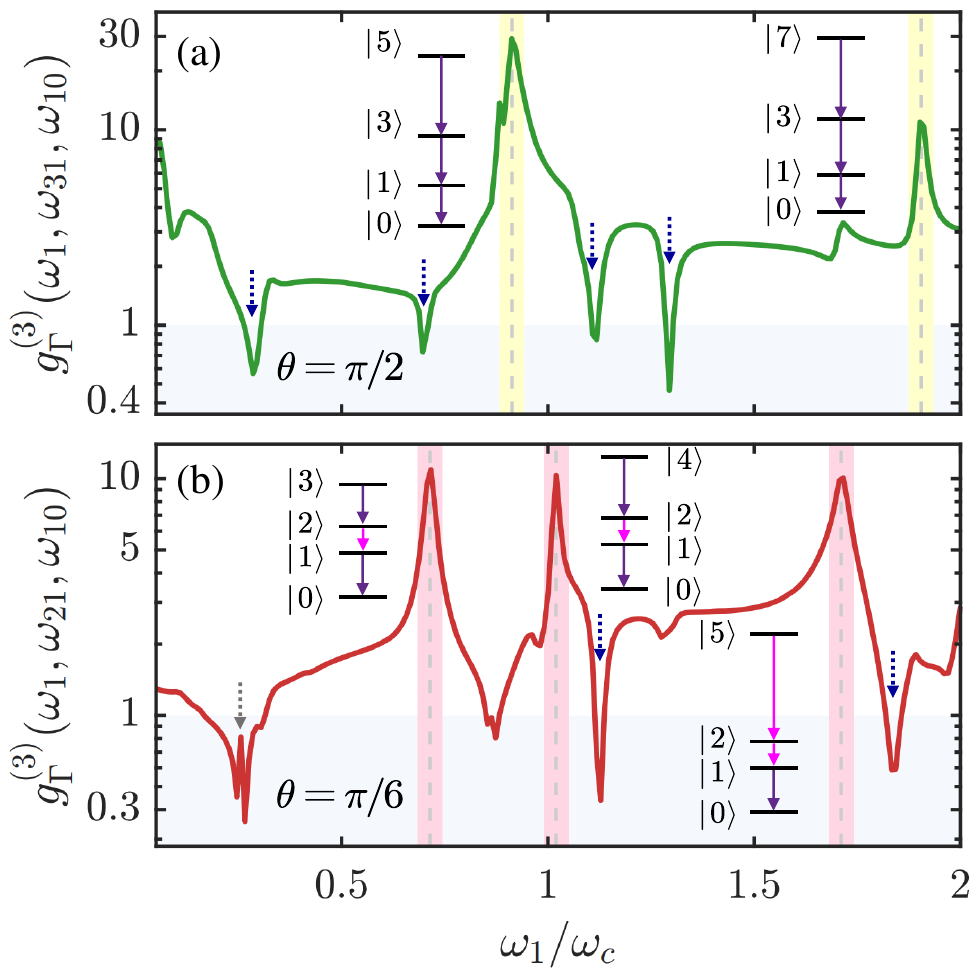}
	\caption{Third-order photon correlation functions $g_{\Gamma}^{(3)}(\omega_1, \omega_{2}, \omega_{3})$ vs sensor resonance frequency $\omega_1$. (a) $\theta=\pi/2$, with two sensors fixed at $\omega_2=\omega_{31}$ and $\omega_3=\omega_{10}$. (b) $\theta=\pi/6$,  with two sensors fixed at $\omega_2=\omega_{21}$ and $\omega_3=\omega_{10}$.
		The light-blue background represents the antibunching regime ($g_{\Gamma}^{(3)}(\omega_1, \omega_{2}, \omega_{3})<1$), whereas the peaks (shaded in yellow and pink) indicate strong photon bunching ($g_{\Gamma}^{(3)}(\omega_1, \omega_{2}, \omega_{3})>1$). Several key dips are marked by blue dotted arrows. The parameters are the same as in Fig.\,\ref{fig:spectrum}.}
	\label{fig:threephoton}
\end{figure}
To investigate frequency-resolved multiphoton correlations, we first analyze the two-photon correlations of the system. Figure\,\ref{fig:twophoton} shows the second-order correlation function $g_{\Gamma}^{(2)}(\omega_1, \omega_2)$ between a photon with fixed frequency $\omega_2=\omega_{10}$ and another photon with frequency $\omega_1$ scanning across the spectra range. Here $\omega_{10}=E_{1}-E_{0}$ corresponds to the transition from first excited state to the ground state.  For the symmetric case $\theta=\pi/2$ shown in Fig.\,\ref{fig:twophoton}(a), the most obvious features are two bunching peaks (yellow shaded regions). These peaks occur at frequencies corresponding to the spectral peaks $\text{D}$ and $\text{I}$ in Fig.\,\ref{fig:spectrum}(b). The physical origin of these correlations can be understood from the cascade emission between dressed eigenstates. Specifically, the cascade transitions from higher excited states to ground state, i.e., $|3\rangle \rightarrow |1\rangle \rightarrow |0\rangle$ and $|4\rangle \rightarrow |1\rangle \rightarrow |0\rangle$, lead to two-photon bunching between photons of different frequencies. In contrast, the antibunching dips arise from disconnected emission pathways. By comparing the positions of the dips (blue arrows) with the spectral peaks, they can be identified with the transitions $|3\rangle \rightarrow |2\rangle$, $|4\rangle \rightarrow |2\rangle$, $|2\rangle \rightarrow |0\rangle$. A common feature of these transitions is that none of them forms a cascade transition with the fixed  $|1\rangle \rightarrow |0\rangle$.

As discussed previously, parity symmetry breaking at $\theta=\pi/6$ activates transition channels that are forbidden in the symmetric case. This symmetry breaking also has a pronounced impact on the photon statistics. As shown in Fig.\,\ref{fig:twophoton}(b), two additional bunching peaks (pink shaded regions) emerge compared with Fig.\,\ref{fig:twophoton}(a). These peaks correspond to the transitions $|2\rangle \rightarrow |1\rangle$ and $|5\rangle \rightarrow |1\rangle$ marked by the magenta arrows, respectively. These newly available transition pathways combine with the reference transition $|1\rangle \rightarrow |0\rangle$ to form cascade processes, resulting in correlated photon pairs with different frequencies. The appearance of these new peaks is accompanied by a slight reduction in the intensity of the original bunching peaks (yellow), reflecting the competition among different radiative pathways. In addition, more antibunching dips appear in the correlation spectrum, as illustrated by the last two dips in the figure. These dips originate from newly allowed transitions such as $|5\rangle \rightarrow |2\rangle$ and $|3\rangle \rightarrow |0\rangle$, which also arise from symmetry breaking. Although the values of these dips remain below unity, they are noticeably larger than in the symmetric case, indicating weak correlations between the corresponding photon pairs. However, since these transitions do not form cascade processes, the emitted photons still exhibit antibunching. These observations naturally motivate the investigation of higher-order photon correlations.

We therefore proceed to study the third-order photon correlation function, $g_{\Gamma}^{(3)}(\omega_1, \omega_{2}, \omega_{3})$, shown in Fig.\,\ref{fig:threephoton}. Unlike the calculation of the second-order correlation $g_{\Gamma}^{(2)}(\omega_1, \omega_{10})$, the sensor frequencies are chosen differently for the third-order correlation $g_{\Gamma}^{(3)}(\omega_1, \omega_{2}, \omega_{3})$ in the cases $\theta=\pi/2$ and $\theta=\pi/6$. For the symmetric case $\theta=\pi/2$, the transition $|2\rangle \rightarrow |1\rangle$ is forbidden by parity selection rules,  we thus fix the two sensors at frequencies $\omega_2=\omega_{31}$ and $\omega_3=\omega_{10}$ in order to observe a potential three-photon cascade process.  For $\theta=\pi/6$, the second excited state $|2\rangle$ can serve as the mediating state. Accordingly, the sensor frequencies are chosen as $\omega_2=\omega_{21}$ and $\omega_3=\omega_{10}$.

In the case $\theta=\pi/2$, as in Fig.\,\ref{fig:threephoton}(a), two strong bunching peaks with $g_{\Gamma}^{(3)}(\omega_1, \omega_{31}, \omega_{10})\gg 1$ appear at 
$\omega_1=\omega_{53}$ and $\omega_1=\omega_{73}$,  corresponding to the cascade processes $|5\rangle \rightarrow |3\rangle \rightarrow |1\rangle \rightarrow |0\rangle$ and $|7\rangle \rightarrow |3\rangle \rightarrow |1\rangle \rightarrow |0\rangle$, respectively, where all involved transitions occur between eigenstates of different parity. Moreover, a striking feature is that the magnitude of these third-order correlation peaks is about one order of magnitude larger than that of their second-order counterparts. This indicates the pronounced frequency-resolved three-photon correlations in the system. Additionally, a smaller sub-peak appears near $\omega_1=\omega_{31}$ close to the main peak at $\omega_{53}$. Because these two transition frequencies are very close, they cannot be fully resolved by the finite spectral resolution of the sensor, resulting in an apparent three-photon bunching signal when $\omega_1=\omega_{31}$. In contrast, when the frequencies of the first photon are $\{\omega_{32}, \omega_{10}, \omega_{42}, \omega_{20}\}$, the joint emission is suppressed, giving rise to antibunching dips. As discussed earlier, these transitions do not form cascade emission pathways. 

For the case $\theta=\pi/6$, the strong frequency-resolved photon bunching ($g_{\Gamma}^{(3)}(\omega_1, \omega_{2}, \omega_{3})>1$) can also be observed, as shown through the pink-shaded peaks in Fig.\,\ref{fig:threephoton}(b), which correspond to processes that involve symmetry-breaking-induced transitions between eigenstates of the same parity. In this case, more three-photon cascade processes become possible, including $|3\rangle \rightarrow |2\rangle \rightarrow |1\rangle \rightarrow |0\rangle$, $|4\rangle \rightarrow |2\rangle \rightarrow |1\rangle \rightarrow |0\rangle$ and $|5\rangle \rightarrow |2\rangle \rightarrow |1\rangle \rightarrow |0\rangle$. These processes give rise to bunching peaks with values $g_{\Gamma}^{(3)}(\omega_1, \omega_{21}, \omega_{10})\approx 10$ (pink shaded regions). The first two cascades involve one symmetry-breaking-induced transition (magenta arrows), while the third involves two such transitions.
Interestingly, these same frequencies correspond to dips in the second-order correlation shown in Fig.\,\ref{fig:twophoton}(b). For instance, the first peak satisfies $g_{\Gamma}^{(3)}(\omega_{32}, \omega_{21}, \omega_{10})>1$ but $g_{\Gamma}^{(2)}(\omega_{32}, \omega_{10})<1$, which means the two photons (with frequency $\omega_{32}$ and $\omega_{10}$) appears antibunched in the two-photon correlation but becomes strongly correlated when the full three-photon cascade is considered. The same behavior occurs for the other two peaks. Compared with the symmetric case, the peak values of the third-order correlation function are reduced. This reduction arises from the competition among multiple cascade pathways.
As a consequence of some dips in the second-order correlation function converting into peaks in the third-order correlation function, the total number of dips decreases. Indeed, only two prominent dips remain at $\omega_1=\omega_{20}$ and $\omega_1=\omega_{30}$. Finally, a notable feature emerges at $\omega_1=\omega_{21}$ (gray arrow). Although this frequency corresponds to a bunching peak in the second-order correlation $g_{\Gamma}^{(2)}(\omega_{21}, \omega_{10})>1$, the third-order correlation satisfies $g_{\Gamma}^{(3)}(\omega_{21}, \omega_{21}, \omega_{10})<1$, indicating antibunching. Physically, once a photon with frequency $\omega_{21}$ is emitted, the system relaxes to the state $|1\rangle$, from which a second identical photon cannot be immediately emitted.

\section{CONCLUSION}\label{sec V}

In conclusion, we have investigated frequency-resolved multiphoton correlations in an incoherently driven cavity QED system operating in the ultrastrong-coupling regime. We derive analytical expressions for the power spectrum, as well as for the frequency-resolved two-photon and three-photon correlation functions. By analyzing the eigenenergies and eigenstates of the system, we show that the power spectrum exhibits multiple peaks whose frequencies precisely match the transition frequencies between dressed eigenstates. This indicates that the emitted radiation consists of photons with different frequency components. Within the dressed-state picture, we first examine two-photon correlations. Both frequency-resolved photon bunching and antibunching are observed at different frequencies, depending on whether the transitions associated with the two photons emission can form a consecutive cascade.  Extending the analysis to higher-order processes, we find that three-photon correlations exhibit significantly stronger bunching, while frequency-resolved three-photon antibunching also appears in certain frequency regions. Furthermore, compared with the symmetry-preserving case, breaking the system symmetry activates additional transition channels and allows more multiphoton cascade processes to occur, thereby enabling the generation of strong bunched $N$-photon emissions ($N=2,3$). Our results may provide useful insights for future studies on multiphoton quantum light sources in the USC regime.

\section*{Acknowledgments}

This work is supported by the National Natural Science Foundation of China (Grants  No.\,12574397 and No.\,12547108), the Quantum Science and Technology-National Science and Technology Major Project (Grant No.\,2024ZD0301000), the National Science Fund for Distinguished Young Scholars of China (Grant No.\,12425502), the National Key Research and Development Program of China (Grant No.\,2021YFA1400700), the Sichuan Science and Technology Program (Grant No.\,2025ZNSFSC0057).
%
	
\end{document}